# Crystal Growth and Characterization of Bulk $Sb_2Te_3$ Topological Insulator

Rabia Sultana[1,2], Ganesh Gurjar[3], S. Patnaik[3] and V.P.S. Awana[1,2*]

[1] National Physical Laboratory (CSIR), Dr. K. S. Krishnan Road, New Delhi-110012, India

[2] Academy of Scientific and Innovative Research (AcSIR), NPL, New Delhi-110012, India

[3] School of Physical Sciences, Jawaharlal Nehru University, New Delhi-110067, India

**Abstract**

The $Sb_2Te_3$ crystals are grown using the conventional self flux method via solid state reaction route, by melting constituent elements (Sb and Te) at high temperature (850°C), followed by slow cooling (2°C/hour). As grown $Sb_2Te_3$ crystals are analysed for various physical properties by X-ray diffraction (XRD), Raman Spectroscopy, Scanning Electron Microscopy (SEM) coupled with Energy Dispersive X-ray Spectroscopy (EDAX) and electrical measurements under magnetic field (6Tesla) down to low temperature (2.5K). The XRD pattern revealed the growth of synthesized $Sb_2Te_3$ sample along (00l) plane, whereas the SEM along with EDAX measurements displayed the layered structure with near stoichiometric composition, without foreign contamination. The Raman scattering studies displayed known ($A_{1g}^1$, $E_g^2$ and $A_{1g}^2$) vibrational modes for the studied $Sb_2Te_3$. The temperature dependent electrical resistivity measurements illustrated the metallic nature of the as grown $Sb_2Te_3$ single crystal. Further, the magneto – transport studies represented linear positive magneto-resistance (MR) reaching up to 80% at 2.5K under an applied field of 6Tesla. The weak anti localization (WAL) related low field (± 2Tesla) magneto-conductance at low temperatures (2.5K and 20K) has been analysed and discussed using the Hikami-Larkin - Nagaoka (HLN) model. Summarily, the short letter reports an easy and versatile method for crystal growth of bulk $Sb_2Te_3$ topological insulator (TI) and its brief physical property characterization.

*Corresponding Author
Dr. V. P. S. Awana:  E-mail: awana@nplindia.org
Ph. +91-11-45609357, Fax-+91-11-45609310
Homepage: awanavps.webs.com

**Introduction**

Nowadays, topological insulators (TIs) are emerging as one of the most promising and challenging research topics in condensed matter physics community, due to their fascinating physical properties and novel device applications such as spintronics, practical quantum computers, optoelectronics and superconductors. [1-10]. TIs are a new class of electronic materials exhibiting both gapped bulk states and gapless Dirac boundary states simultaneously [10-12]. Interestingly, in comparison to other materials, TIs have gained special attention not only due to their tremendous potential for various applications but also due to relatively low cost. The existence of narrow band gap, protected conducting surface edges states, formation of single mass-less Dirac cone and strong optical absorbance along with low saturating optical intensity are some of their unique properties [1,8,7,11]. The presence of unique topological quantum states is confined by spin orbit interaction and time reversal symmetry, which prevents backscattering from impurities and defects [1-12]. In TIs the electrons are localized in the bulk but delocalized at the edges [13]. Furthermore, TIs behaves as excellent thermoelectric (TE) materials due to the presence of analogous material features such as heavy elements and narrow band gaps [12,13].

Similar to $Bi_2Se_3$ and $Bi_2Te_3$, $Sb_2Te_3$ is a well-studied and experimentally confirmed typical 3D layered TI exhibiting a relatively large bulk energy gap and a single surface Dirac cone [11, 15-17]. In case of TIs, it is also interesting to note that a layered material exhibits rhombohedral structure with in $R\bar{3}m$ (D5) space group. Apart from being a known TI and TE, the $Sb_2Te_3$ holds great promise as a phase changing material for information storage in chalcogenide alloys [18]. Various synthesis techniques; such as Bridgman crystal growth, chemical vapour deposition (CVD), pulsed-laser deposition (PLD), flux free method, metal organic chemical vapour deposition (MOCVD), and molecular beam epitaxy (MBE) have been employed to grow bulk single crystals, thin films, nano-plates/nano-ribbons and nano-sheets of TI and topological superconductor (TSC) materials [6, 17,19, 26-37]. Interestingly, there are relatively less reports on the synthesis of bulk singly crystalline TI materials, using the flux free method. In this direction, we recently reported the flux free growth of $Bi_2Te_3$ and $Bi_2Se_3$ single crystals by solid state reaction route from their high temperature melt and slow cooling [34, 35]. The precise hold time for the liquid state and the cooling rate for the growth of bulk 3D TIs is important. Worth mentioning is the fact that the studied $Sb_2Te_3$ crystal is grown by simple and versatile easily adaptable method without usage of specialized Bridgeman technique and is of reasonable good quality as its host of measured physical properties are same to the ones reported earlier. Summarily, in current letter we report an easy and versatile method for crystal growth of bulk $Sb_2Te_3$ topological insulator (TI) along with its brief physical property characterization.

**Experimental Details**

The bulk $Sb_2Te_3$ single crystal was grown by the conventional solid state reaction route via self flux method. Though the growth method is similar to our previous works on $Bi_2Te_3$ and $Bi_2Se_3$ [34,35], the heating schedule for presently synthesized $Sb_2Te_3$ crystal is slightly different. The stoichiometric amounts of the constituent elements were accurately weighed, well mixed and grounded thoroughly inside a high-purity argon atmosphere glove box (MBRAUN Lab star) and the powder was then pelletized into rectangular form and sealed in an evacuated quartz tube ($10^{-3}$ Torr) in order to avoid oxidation of the sample during heat treatment. The vacuum encapsulated quartz tube was heated to 850°C with a heating rate of 2°C min$^{-1}$ and hold for 12 hours, cooled slowly at a rate of 2°C hour$^{-1}$ to 650°C, and finally the furnace was switched off to cool down. The details were reported by some of us recently in a conference paper [38]. Inset of Figure 1 shows the schematic diagram of heat treatment of the synthesized bulk $Sb_2Te_3$ single crystal. The as grown $Sb_2Te_3$ single crystal resulted in one piece (~1cm in size) and exhibited shining surfaces [inset of Figure 2].

The structural analysis of as grown $Sb_2Te_3$ single crystal was determined from the room temperature x-ray diffraction (XRD) patterns recorded using Rigaku Miniflex II powder x-ray diffractometer with CuKα radiation ($\lambda$=1.54 Å). Raman spectroscopy measurements were carried out at room temperature using a Renishaw Raman Spectrometer to determine the vibrational properties of the studied $Sb_2Te_3$ single crystal. The morphological analysis was performed by a ZEISS-EVO MA-10 scanning electron microscope (SEM), and energy dispersive X-ray was employed to confirm the stoichiometry of $Sb_2Te_3$. The electrical resistivity measurements under magnetic field are carried out on a Cryogenics system.

**Results and Discussion**

The crystalline phase structure of the synthesized $Sb_2Te_3$ crystal was analyzed by the x-ray diffraction (XRD) at room temperature. Figure 1 shows the XRD pattern on surface of as synthesized $Sb_2Te_3$ crystal. It is clear from Figure 1, that all the diffraction planes are aligned in c-direction i.e., (001) with l = 3, 6, 9, 12, 15, 18, 21, 24. Clearly, the surface XRD analysis of $Sb_2Te_3$ reveals that the diffraction peaks are well defined along c-axis indicating its single crystalline nature.

In order to check the crystal quality i.e., existence of any foreign phases or impurities, powder XRD pattern was taken on gently crushed crystal. Figure 2 displays Rietveld fitted powder XRD pattern of representative $Sb_2Te_3$ crystal, performed using the Full Prof Suite toolbar software. The lattice parameters obtained after fitting area = b = 4.267(2)Å and c = 30.468(5)Å. These values are in agreement to the earlier reported literature [18]. Similar to $Bi_2Te_3$ and $Bi_2Se_3$, $Sb_2Te_3$ shares the identical rhombohedral structure fitted in $R\bar{3}m$ (D5) space group [34, 35, 39-40]. Further, the XRD patterns of the resultant $Sb_2Te_3$ crystal showed no impurity phase, implying its purity within the XRD limits.

Figure 3 shows the schematic unit cell structure of as grown $Sb_2Te_3$ single crystal produced using the VESTA software. The atomic arrangement of the resultant bulk $Sb_2Te_3$ single crystal displays a layered rhombohedral structure consisting of quintuple layers (QLs). Here, each quintuple layer is seen to be comprised of five mono-atomic planes of Te–Sb–Te–Sb–Te. Besides, QLs are bonded by van der Waals (vdW) forces, while the five atoms in each QL are held simultaneously by covalent bond. This result matches well with the previously reported literature [6, 39, 41].

To understand the phonon dynamics, crystalline phase structure, and stoichiometry of $Sb_2Te_3$ single crystal more accurately, we performed Raman spectroscopy measurements. K.M.F Shahil et al. have reported the micro-Raman spectroscopy study of the thin crystalline films of $Sb_2Te_3$ [42], whereas Y. Kim et al. have reported the temperature dependence of $A_{1g}^2$ phonon mode alone in $Sb_2Te_3$ single crystal [43]. On the other hand P. Srivastava et al., reported only $E_g^2$ mode for layered $Sb_2Te_3$ nano-flakes [44]. Also, J. E Boschker et al. showed characteristic Raman peaks in $Sb_2Te_3$ films grown on monolayer graphene [45], whereas S. M. Souza et al. reported the Raman spectra for nano-metric $Sb_2Te_3$ as a function of pressure [46]. However, to the best of our knowledge Raman spectroscopy measurements on bulk $Sb_2Te_3$ single crystal in expanded range for all the three vibrational modes have not been reported so far. Here, we report the room temperature Raman spectra of the as synthesized bulk $Sb_2Te_3$ single crystal recorded using the Renishaw Raman Spectrometer [Figure 4(a)]. The vibrational modes at room temperature were observed using spectral resolution of $0.5 cm^{-1}$ and 2400 I/mm grating. Here, laser source having an excitation photon energy of 2.4eV (514nm) was employed. Moreover, the Laser power applied at the surface of the sample was less than 3mW in order to prevent the sample from damage. The measured spectral range is 10–400cm$^{-1}$, representing three main distinct Raman active vibrational peaks namely $A_{1g}^1$, $E_g^2$ and $A_{1g}^2$. The obtained values for the vibrational modes corresponding to $A_{1g}^1$, $E_g^2$ and $A_{1g}^2$ are 68.77, 112.32 and 165.30cm$^{-1}$ respectively. When compared to the earlier reported TIs by some of us [34, 35], these Raman active vibrational modes positions lies between those as obtained for $Bi_2Te_3$ and $Bi_2Se_3$. Apparently, Raman shift of the corresponding $A_{g1}^1$, $Eg^2$ and $A_{1g}^2$ vibrational modes increases, where as the peak intensities were observed to be decreasing simultaneously. Further, the observed values matches well and are in good agreement with the one being theoretically calculated for $Sb_2Te_3$ [18]. To facilitate a direct comparison of the three popular TIs i.e., $Sb_2Te_3$, $Bi_2Se_3$ and $Bi_2Te_3$, Figure 4(a), (b) and (c) respectively depict their experimentally observed Raman spectrums at room temperature. The data for $Bi_2Te_3$ and $Bi_2Se_3$ are taken from our previous work ref. 34 and 35. Apparently, Figure 4 shows that $Bi_2Se_3$ TI exhibits higher values of Raman shift as well as intensity followed by $Sb_2Te_3$ and $Bi_2Te_3$. The existence of both higher Raman shift values as well as intensity in case of $Bi_2Se_3$ may result due to the stronger bonding forces in comparison to $Sb_2Te_3$ and $Bi_2Te_3$ TIs, along with lighter atomic weight of Se as compared to the Te atom [42]. However, the $Sb_2Te_3$ TI exhibits Raman spectra nature similar to $Bi_2Se_3$. The reason is that Te is heavier than Se on the other hand Sb is lighter than Bi and hence the $Sb_2Te_3$ and $Bi_2Se_3$ are quite similar to each other and different than $Bi_2Te_3$. These are only qualitative conjectures and more precise Raman studies could further quantify the situation.

Figure 5(a) illustrates the typical SEM images of $Sb_2Te_3$ single crystal taken with a scale of 2μm. The characteristic SEM image displayed a clean, smooth and layered slab like structure indicating good quality of the synthesized crystal. The slab like structure is aligned in c- direction. The uni-directional parallel planes being seen in Figure 5(a), confirms the one plane (00l) orientation of the studied $Sb_2Te_3$ single crystal. Further, to investigate the purity and stoichiometry of as grown $Sb_2Te_3$ single crystal, EDAX measurements were done, which are shown in Figure 5(b). The compositional constituents of as grown $Sb_2Te_3$ crystal [Figure 5(b)] revealed the existence of only Sb and Te atoms, without any impurity contamination, indicating the purity of the synthesized crystal. The EDAX result obtained are in agreement with the XRD results [Figure 2]. The inset of Figure 5(b) depicts the quantitative weight % values of Sb (40%) and Te (60%) atoms respectively. Consequently, the EDAX analysis showed uniform chemical composition and confirmed the 2:3 ratio of Sb and Te atoms, which are close to the stoichiometric $Sb_2Te_3$ [inset of Figure 5(b)]. Summarily, the detailed

SEM and EDAX analysis showed the unidirectional growth and near perfect stoichiometry of the studied $Sb_2Te_3$ crystal.

The electrical resistivity of as synthesized $Sb_2Te_3$ single crystal as a function of temperature from 250K down to 2.5K is shown in Figure 6(a). The electrical resistivity is observed to decrease with decrease in temperature indicating a metallic behaviour from 250 to 2.5K. Inset in Figure 6(a) shows the hallmark of Fermi liquid behaviour of $Sb_2Te_3$ single crystal as being fitted using $\rho=\rho_0+AT^2$, where $\rho$, $\rho_0$, A and T represents the resistivity, residual resistivity, constant and the temperature respectively. The resistivity curve clearly obeys the Fermi liquid behaviour in a temperature range from 5K to 50K. Here, the residual resistivity ($\rho_0$) defined as the resistivity at zero temperature (0K) is obtained by extrapolation as 0.0096mΩ-cm. On the contrary, the value of resistivity at 250K ($\rho_{250}$) is found to be 0.0685mΩ-cm, accordingly, the residual resistivity ratio ($\rho_{250}/\rho_0$) value comes out to be 7.1354.

Figure 6 (b) displays the temperature dependent electrical resistivity for $Sb_2Te_3$ single crystal under different applied magnetic fields in temperature range of 2.5 to 100K. As the magnetic field is increased from 0 to 5Tesla the resistivity value increases considerably with increase in applied field. Also the resistivity curves under different applied fields displays metallic behavior analogous to that observed under the absence of magnetic field [Fig. 6(a)]. Additionally, one can clearly observe that the resistivity curve tends to saturate at lower temperatures, say below 20K. This tendency of saturation may possibly occur from surface conduction as well as bulk residual conduction as reported earlier in $Bi_2Se_3$ topological insulator [24].

Figure 7(a) displays the MR data obtained under applied magnetic fields of up to 6Tesla at various temperatures for the studied $Sb_2Te_3$ single crystal. The MR value was obtained using the formula MR(%) = {[$\rho(H) - \rho(0)$] / $\rho(0)$}*100, where $\rho(H)$ and $\rho(0)$ are the resistivity with and without applied magnetic field (H) respectively. Accordingly the calculated MR(%) value reaches up to 80% as seen in Figure 7(a) and the shape of the plot is nearly V type at 2.5K under 6Tesla magnetic field. However, this linear positive nature of MR curve gradually tends to broaden at higher temperatures (up to 250K) and takes a U type shape with much decreased MR% values. Quantitatively, the MR values decrease from 80% to 12% with increase in temperature from 2.5K to 250K under applied fields up to 6Tesla.

It is interesting to note that the MR curve at 2.5K displayed a typical v- type cusp (sharp dip) with in the overall V shape at lower magnetic fields i.e., below ±2Tesla. This typical behavior at low temperature and magnetic field is popularly known as WAL (weak anti localization) effect [47-52]. Further, this behavior is observed to disappear gradually with increase in temperature and magnetic field due to the decrease in the phase coherence length as to be discussed below. As far as the value of linear non saturating MR% is concerned the same is much less (80%) in comparison to 250% at 2.5K under 6Tesla for studied $Sb_2Te_3$ and our earlier reported $Bi_2Te_3$ [34]. The MR% values may differ for various studied TI single crystals, depending upon the contributions from surface and bulk states, as well the possible abundance of defects or vacancies.

Figure 7(b) shows the WAL related low field magneto- conductance curve of $Sb_2Te_3$ single crystal fitted using the well – known Hikami - Larkin - Nagaoka (HLN) model [53]:

$$\Delta G(H) = -\frac{\propto e^2}{\pi h}\left[\ln(\frac{B_\varphi}{H}) - \Psi\left(\frac{1}{2} + \frac{B_\varphi}{H}\right)\right]$$

Where, $\Delta G(H) = G(H) - G(0)$ represents the change of magneto-conductance, α is the WAL coefficient indicating the strength of the WAL, e denotes the electronic charge, h represents the Planck's constant, Ψ is the digamma function, H is the applied magnetic field, $B_\varphi = \frac{h}{8e\pi H l_\varphi}$ is the characteristic magnetic field and $l_\varphi$ is the phase coherence length. The magneto-conductance data was fitted using HLN model at lower magnetic fields up to (± 2Tesla) and at lower temperatures (2.5K and 20K) respectively. Here, α and $l_\varphi$ are the fitting parameters. The value of α at 2.5 K, extracted using the HLN model is -0.06 whereas, the $l_\varphi$ value calculated is 41nm. At 20K, the value of α and $l_\varphi$ comes out to be -0.22 and 15nm respectively. Accordingly, as the temperature increases from 2.5K to 20K the value of both α and $l_\varphi$ appears to decrease which is in agreement with other TI systems [49, 51, 54]. The inset of Figure 7(b) shows the zoom in view of the HLN fitted curves (2.5K and 20K) at low fields up to ± 2Tesla.The value of α and $l_\varphi$ varies from sample to sample including bulk single crystals, thin films, nano wires and thin flakes [47-52, 54,55]. Ideally, the value of α, determining the number of coherent transport channels should be -0.5 for single coherent channel and between -0.5 to -1.5 for multi parallel conduction channels (surface and bulk states) [47-55]. However, the experimentally fitted values of α is reported to lie between –0.4 and –1.1, for single surface state, two surface states , or intermixing between the surface and bulk states [50, 55]. In fact, depending upon the relative contribution of WAL from the surface channels and weak localization (WL) of the bulk channels, the value of α may differ considerably [55]. In our case, the low temperature (2.5K and 20K) and low field (± 2Tesla) the values of α is quite low, indicating very thin conducting surface, which is an indication of the comparatively more contribution from bulk (WL) than the conducting surface states (WAL). Further, the bulk part can induce WL in addition to WAL, indicating that the small value of α may be the mixing of both WL and WAL [55].This is clear from the fact that in our case, the coherence length ($l_\varphi$) is also relatively lower (41nm and 15nm) at 2.5K and 20K in comparison to some other reports [47-52, 54]. The studied $Sb_2Te_3$ single crystal follows the HLN model at low temperatures and fields, albeit with thin conducting surface and relatively lower coherence length. This is occurring due to the competing WAL (surface) and bulk induced WL states in the studied $Sb_2Te_3$ single crystal.

**Conclusion**

We conclude to have successfully synthesized bulk $Sb_2Te_3$ topological insulator via the self flux method having single crystalline nature with 00l alignment. The unit cell displayed the rhombohedral structure consisting of quintuple layers, whereas the Raman measurements showed theoretically predicted vibrational modes present in the crystal. The SEM and EDAX analysis confirmed the layered crystalline nature, purity and stoichiometry of the resultant $Sb_2Te_3$ crystal. The electrical resistivity measurements showed the metallic nature as well as Fermi liquid behavior at lower temperatures below say 50K. The MR measurements displayed positive linear value reaching up to 80% at 2.5K under 6Tesla magnetic field, which could be useful in various potential applications. Further the studied $Sb_2Te_3$ single crystal data are fitted in HLN model at low temperatures and fields, resulting in thin conducting surface and relatively lower coherence length due to competing surface (WAL) and bulk (WL) states.

**Acknowledgements**

The authors from CSIR-NPL would like to thank their Director NPL, India, for his keen interest in the present work. Authors further thank Dr.Bhasker Gahtori for SEM and Mrs. Shaveta Sharma for Raman studies. S. Patnaik thanks DST-SERB project (EMR/2016/003998) for the low temperature high magnetic facility at JNU, New Delhi.

Rabia Sultana and Ganesh Gurjar thank CSIR, India, for research fellowship. Rabia Sultana thanks AcSIR-NPL for Ph.D. registration.

**Figure Captions**

**Figure 1:** X-ray diffraction pattern of as synthesized $Sb_2Te_3$ single crystal. Inset shows the schematic heat treatment diagram of $Sb_2Te_3$ single crystal.

**Figure 2:** Rietveld fitted room temperature XRD pattern for powder $Sb_2Te_3$ crystal. Inset displays the image of as grown $Sb_2Te_3$ single crystal.

**Figure 3:** Unit cell structure of $Sb_2Te_3$ single crystal.

**Figure 4:** Raman spectra for bulk (a) $Sb_2Te_3$ (b) $Bi_2Se_3$ and (c) $Bi_2Te_3$ single crystal at room temperature.

**Figure 5 (a):** Scanning electron microscopy image for $Sb_2Te_3$ single crystal, showing aligned planes in c-direction. .

**Figure 5 (b):** Compositional constituents of the studied $Sb_2Te_3$ single crystal. Inset represents the quantitative weight% values of the atomic constituents of $Sb_2Te_3$ (Sb and Te).

**Figure 6 (a):** Temperature dependent electrical resistivity of $Sb_2Te_3$ in a temperature range of 250K to 2.5K. Inset red solid curve shows Fermi-liquid behavior fit using equation $\rho=\rho_o+AT^2$.

**Figure 6 (b):** Temperature dependent electrical resistivity for $Sb_2Te_3$ single crystal under different applied magnetic field.

**Figure 7 (a):** MR (%) as a function of magnetic field (H) for $Sb_2Te_3$ single crystal at different temperatures.

**Figure 7 (b):** WAL related magneto-conductance for $Sb_2Te_3$ single crystal at 2.5K and 20K fitted using the HLN equation. Inset shows the zoom in view of the fitted curve at low fields up to ± 2Tesla.

Fig. 1

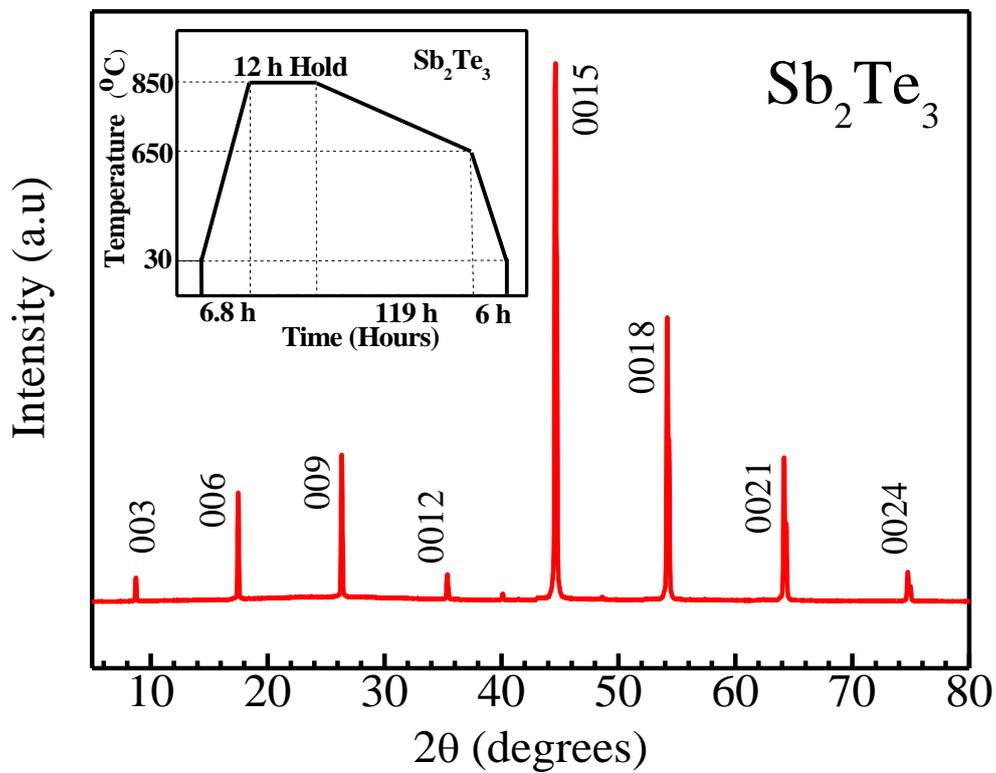

Fig. 2

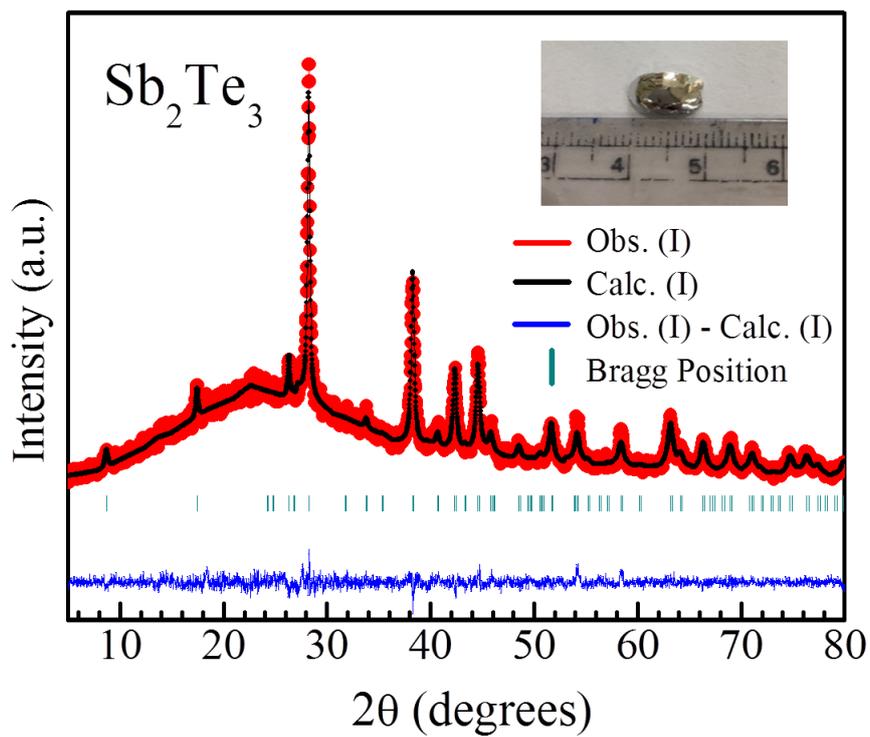

Fig. 3

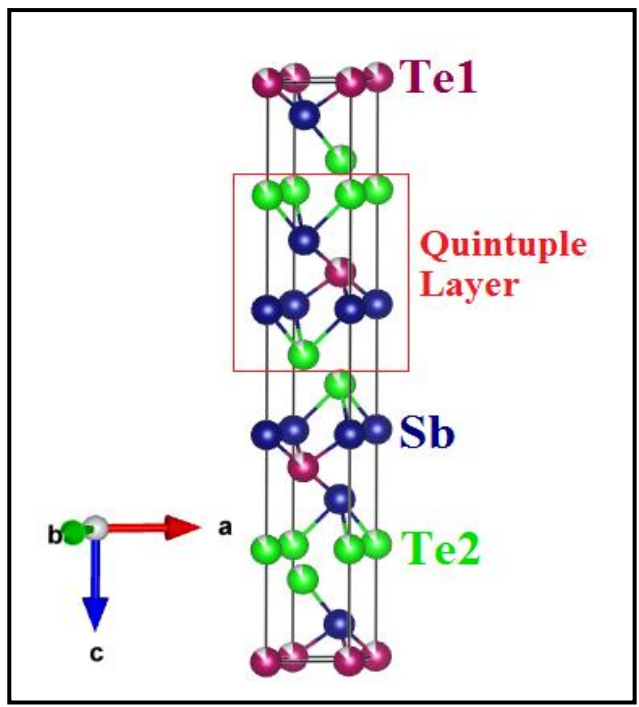

Fig. 4

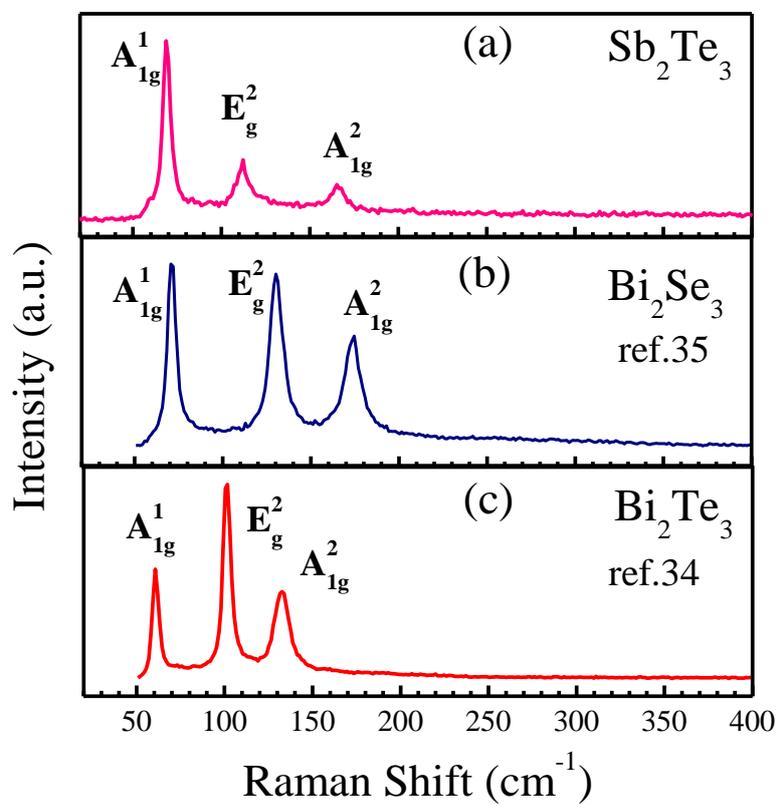

Fig. 5(a)

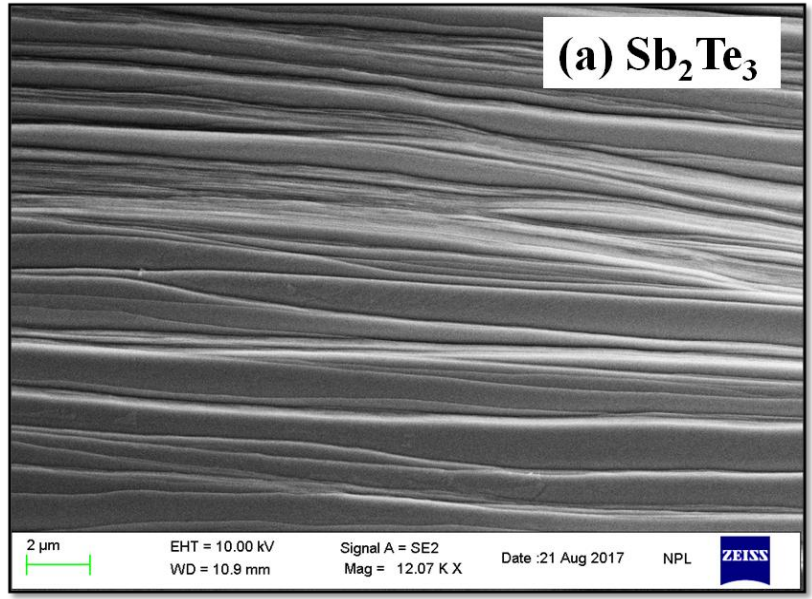

Fig. 5(b)

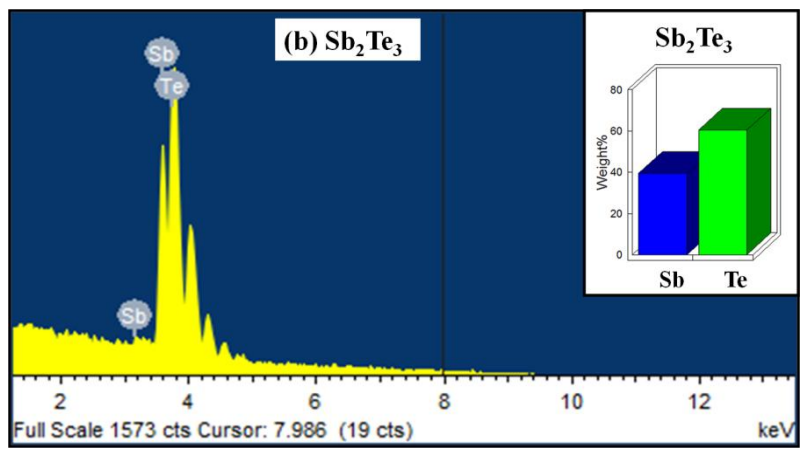

Fig. 6(a)

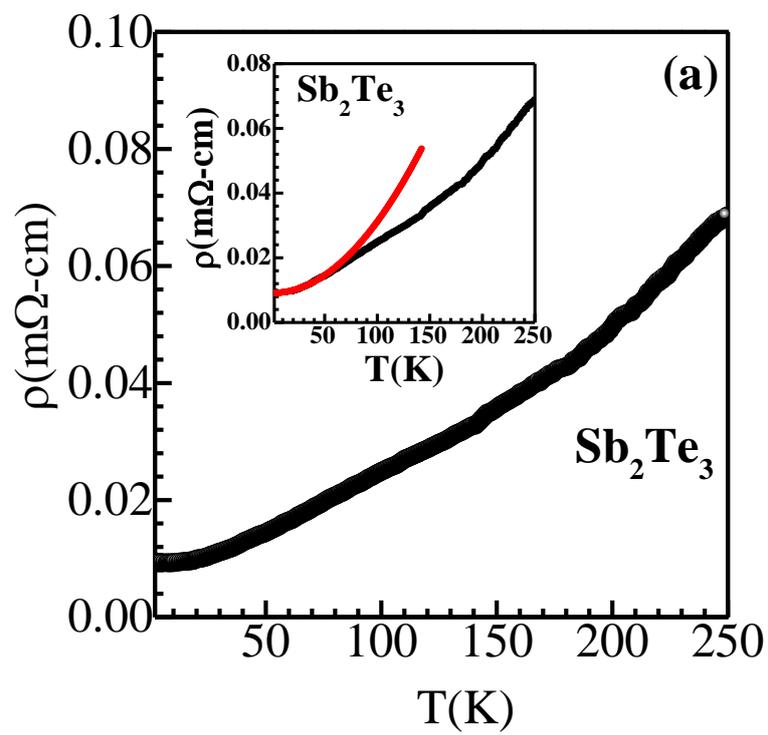

Figure 6(b)

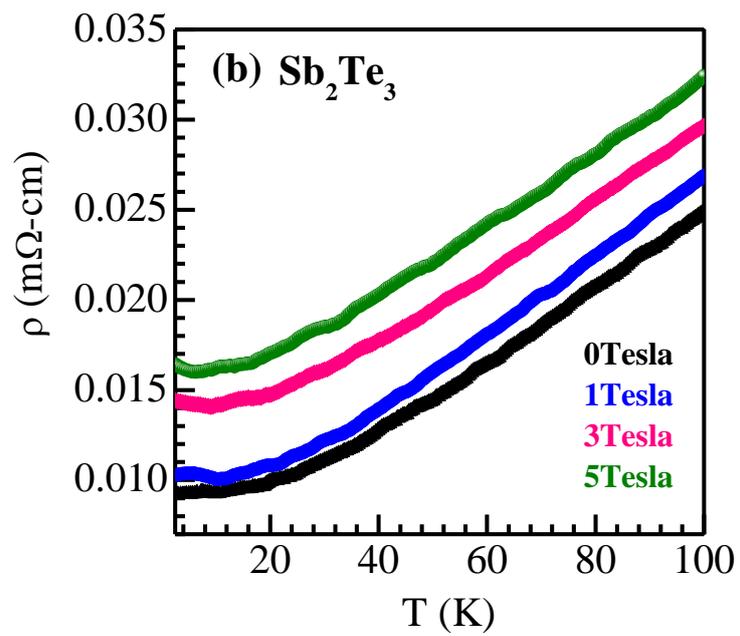

Fig. 7(a)

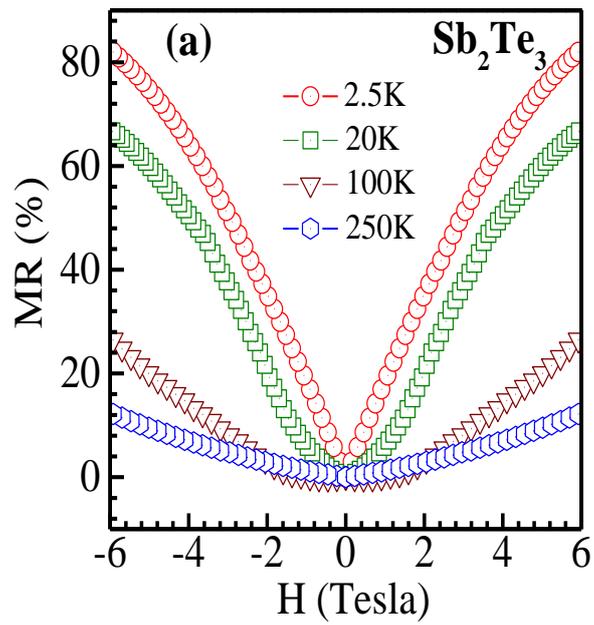

Fig. 7(b)

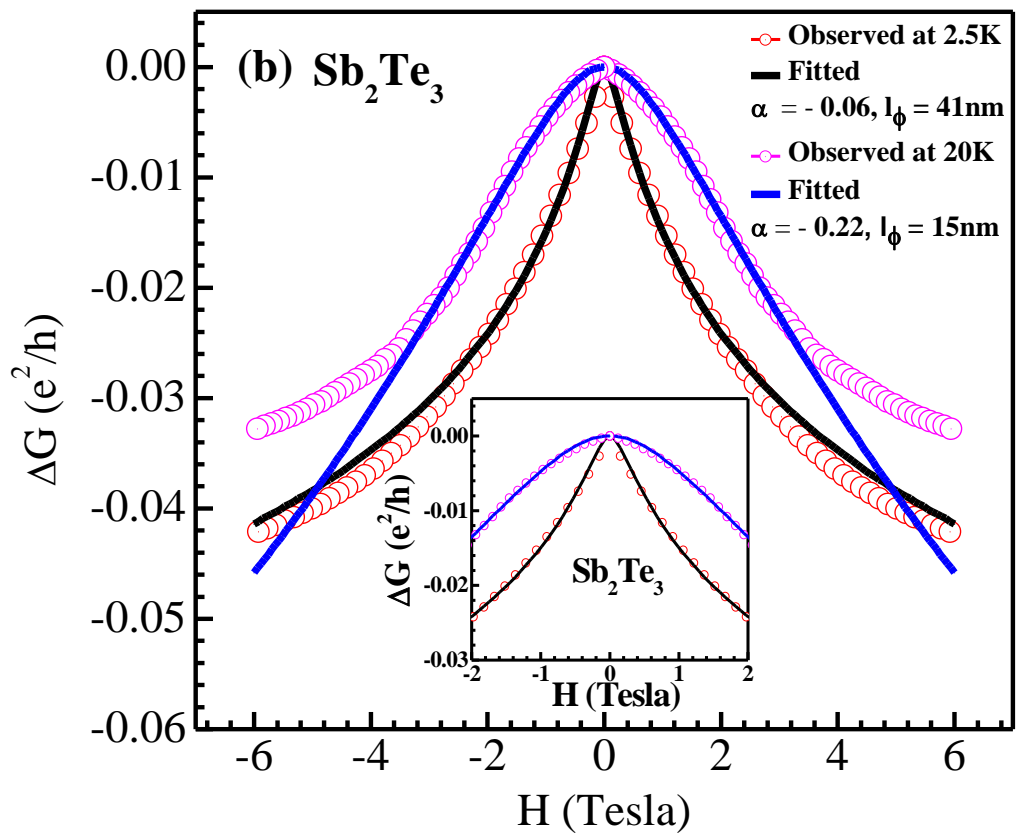